\documentclass{ws-procs9x6}
\usepackage{amsmath}
\usepackage{amssymb}
\usepackage{amsfonts}
\usepackage{latexsym}
\begin{document}

\title{Topological Quantum Field Theory\\
and\\
Pure Yang-Mills Dynamics}

\author{Jan GOVAERTS}

\address{Institute of Nuclear Physics, Catholic University of Louvain\\
2, Chemin du Cyclotron, B-1348 Louvain-la-Neuve, Belgium\\
E-mail: govaerts@fynu.ucl.ac.be}


\maketitle

\abstracts{
By considering specific limits in the gauge coupling constant of
pure Yang--Mills dynamics, it is shown how there exist topological quantum
field theory sectors in such systems defining nonperturbative topological
configurations of the gauge fields which could well play a vital role
in the confinement and chiral symmetry breaking phenomena of
phenomenologically realistic theories such as quantum chromodynamics,
the theory for the strong interactions of quarks and gluons.
A general research programme along such lines is outlined.
A series of other topics of possible relevance, ranging from
particle phenomenology to the quest for the ultimate unification 
of all interactions and matter including
quantum gravity, are raised in passing. The general discussion is
illustrated through some simple examples in 0+1 and 1+1 dimensions,
clearly showing the importance of properly accounting for quantum
topological properties of the physical and configuration spaces
in gauge theories based on compact gauge groups.
\vspace{-10pt}
}

\section{Introduction}
\label{Sec1}

Topological field theories\cite{Witten1.1,Witten1.2,Witten2,TQFT}
are dynamical systems possessing local
or gauge invariances so large and powerful that their physical or
gauge invariant sector solely depends on the topology --- more
precisely, the differentiable to\-po\-lo\-gy class --- of the space
on which these theories are defined. In other words, gauge invariant
states and observables in such systems determine topological invariants
of the underlying manifold. Once quantised, these features survive,
possibly modulo some global aspects related to quantum anomalies.\cite{Witten2}
Such topological quantum field theories (TQFT's) have become of great
interest in the past fifteen years,\cite{TQFT} and are called to play 
a vital role in a variety of fields in pure mathematics, and in mathematical 
and theo\-re\-ti\-cal physics, including the quest for a fundamental 
quantum unification of all particles and interactions. Even in the case 
of quantum mechanical systems --- of which the degrees of freedom 
are function of time only ---, there exist examples of such 
topological systems of interest. However,
it is in the context of gauge field theories, whether of the Yang--Mills
or gravity type, that TQFT's have proved to be of great value. 

Among TQFT's, one distinguishes\cite{TQFT} the so-called theories of
Schwarz type and of Witten type. Theories of Schwarz type
may be defined through a local action principle independent of a
metric structure on the underlying manifold. Theories of Witten type
require such a metric structure to be specified in order to define
their dynamics through an action principle, but their gauge symmetries 
are such that the physical or gauge invariant observables are 
independent of the metric structure nonetheless. A favourite example 
of a TQFT of Schwarz type is 2+1-dimensional pure Chern-Simons theory, 
whether for an abelian or nonabelian gauge group.\cite{Witten2,GovCS} 
A favourite example of a TQFT of Witten type is Witten's original construction
related to Donaldson's topological classification of 4-dimensional
differential structures based on nonabelian (anti)self-dual instanton
solutions to the Yang--Mills equations.\cite{Witten1.1} Another reason 
for the fascination with TQFT is that, for instance, pure quantum gravity 
in 2+1 dimensions is, as a matter fact, a Chern-Simons theory based 
on a noncompact gauge group,\cite{Witten3} possibly hinting at a deeper 
connection with a fundamental unification of all interactions including 
quantum gravity.\cite{Witten1.1,Witten1.2}

Another source of challenging open problems is that of confinement
and dynamical chiral symmetry breaking
in quantum nonabelian Yang--Mills theories. Even though all the available
evidence\cite{vanBaal} concurs with the expectation that in such theories, 
indeed quarks and gluons condense and
remain confined into massive colourless bound states, whatever the
excitation energies applied onto these systems, there is no clear and
definite understanding of the actual dynamics responsible for this
feature quite unique to this large class of gauge theories, of direct
relevance to the world of elementary particles and their strong interactions.
Many a suggestion has been made,\cite{Wilson} involving either monopole- or
instanton-like configurations,\cite{Mandelstam,tHooft,Polyakov} 
but the verdict is still to be reached
as to the actual culprits responsible for the confinement
and chiral symmetry breaking phenomena. However, monopoles
and instantons are specific gauge field configurations
pos\-ses\-sing non\-tri\-vial topological properties in configuration space
made manifest through a nontrivial topology in space(time).

Consequently, with the discovery of TQFT's, the suggestion arises that
the dynamics responsible for confinement and chiral symmetry breaking
in Yang--Mills theories could possibly be dominated by a purely 
topological sector of Yang--Mills configuration space, with further 
corrections induced by topologically trivial quantum fluctuations 
around the topological nonperturbative sectors to be accounted for 
in a complete dynamical quantum treatment. The purpose of the present 
contribution is to try define a programme into that direction, 
by pointing out a series of puzzling properties of pure Yang--Mills 
theories such that in specific limits of the gauge coupling constant, 
one ends up with purely TQFT descriptions. It appears that indeed, 
there are TQFT's at work behind the scenes of the complete Yang--Mills 
dynamics. What could the physical consequences of these TQFT sectors be?
Do they have any bearing on the confinement and chiral symmetry
breaking issues?

The purpose of this contribution is foremost to outline a possible
research programme along such lines, raising as we proceed some other
questions reaching beyond that specific goal, hoping to
sufficiently entice some of the younger participants to the Workshop
to launch their own research work into any such direction. Only a few
simple illustrative examples are also presented, even though much more
could be said already at this stage. In Sec.~\ref{Sec2}, the general
ideas of the programme are described. Section~\ref{Sec3} briefly touches
onto the possibility of defining generalised Yang--Mills dynamics
and their eventual physics interest. Then Sec.~\ref{Sec4} presents the
simplest illustration of our discussion in 0+1 dimensions, while
in Sec.~\ref{Sec5} pure abelian U(1) gauge theory in 1+1 dimensions 
is explicitly solved in the spirit of this contribution,
to conclude with some comments in Sec.~\ref{Sec6}.

A word of apology may be in order. Given the purpose with which this
contribution is written, no serious attempt has been made to provide
a comprehensive list of references to the literature in which topics
are studied that are also related to the present discussion. Only a few,
and hopefully useful ``entry points'' are provided. We apologize to any author
who should feel that his/her contribution has been unduly left unmentioned.

\vspace{-7pt}

\section{Pure Yang--Mills Dynamics}
\label{Sec2}

\subsection{The Action Principle}
\label{Sec2.1}

For the sake of the argument, let us consider flat 4-dimensional Minkowski
spacetime (and more generally a $D$-dimensional spacetime)
with a metric of signature $\eta_{\mu\nu}={\rm diag}\,(-+++)$
($\mu=0,1,\cdots,D-1$ and $x^\mu=(ct,\vec{x}\,)$), 
even though most of the discussion hereafter 
readily extends to a space of arbitrary topology and geometry. Furthermore, 
in order to bring to the fore important topological properties, let us 
also compactify the space coordinates into a $d$-torus $T_d$, with $d=D-1$, 
giving spacetime the topology $\mathbb{R}\times T_d$, keeping in mind that at 
the very end a decompactification limit must be applied to final results. 
Even though such a compactification breaks manifest full Poincar\'e covariance 
at intermediate stages, the purpose of such a procedure is twofold. 
On the one hand, since quantisation will be considered following the 
Hamiltonian operator quantisation path,\cite{GovBook,Gov1} 
full Poincar\'e covariance is not manifest anyway. And on the other hand, 
such a compactification renders the normalisation issue of quantum states 
and operators better behaved, momentum eigenstates being then labelled by 
a discrete, rather than a continuous set of indices, while infra-red 
divergences associated to massless excitations of fields are then avoided 
altogether. The choice of a torus topology is made for ease of calculation, 
since space(time) then remains a flat manifold invariant under translations, 
even though the full Lorentz group is then broken down entirely or, at best,
to a discrete finite rotation subgroup depending on the shape of the
spatial torus $T_d$.

Consider now an arbitrary simple compact Lie group $G$ with hermitian
generators $T^a$ and Lie algebra $[T^a,T^b]=if^{abc}T^c$, $f^{abc}$ being the
corresponding structure constants. As is well-known,\cite{Gov1} associated 
to this algebraic structure, one may introduce a Yang--Mills (YM) field 
$A^a_\mu$ ($\mu=0,1,\cdots,d$) and its field strength
$F^a_{\mu\nu}=
\partial_\mu A^a_\nu-\partial_\nu A^a_\mu-f^{abc}A^b_\mu A^c_\nu$,
both varying under spacetime dependent local gauge transformations such that
\begin{equation}
\begin{array}{r c l}
{A^a_\mu}'(x)T^a&=&U(x)A^a_\mu(x)T^aU^{-1}(x)\,+\,i
\partial_\mu U(x) U^{-1}(x)\ ,\\
 & & \\
{F^a_{\mu\nu}}'(x)T^a&=&U(x) F^a_{\mu\nu}(x)T^a U^{-1}(x)\ ,
\end{array}
\end{equation}
where $U(x)=e^{i\theta^a(x)T^a}$ are arbitrary $G$-valued continuous 
functions of the spacetime coordinate $x^\mu=(ct,\vec{x}\,)$.

A gauge invariant dynamics of this pure YM theory is defined
by the Lagrangian density,\footnote{Note that in this context, it
is only the parameter $e^2$ which is relevant, so that the sign of
the gauge coupling constant $e$ is irrelevant for a pure Yang--Mills theory.}
\begin{equation}
\mathcal{L}^{D=4}_{\rm YM}=-\frac{1}{4e^2}F^a_{\mu\nu}F^{a\mu\nu}\,+\,
\frac{1}{4}\theta\epsilon^{\mu\nu\rho\sigma}F^a_{\mu\nu}F^a_{\rho\sigma}\ .
\label{eq:YM4}
\end{equation}
Note well that in the r.h.s. of this expression, the first contribution
requires a spacetime metric structure to raise and lower the spacetime indices
$\mu$ and $\nu$, whereas the second contribution is specific to $D=4$
dimensions, is independent of any metric structure over spacetime, and
is topological in character, corresponding to a Pontryagin/Chern class. 
As a matter of fact, this last term does not contribute
at all to the classical equations of motion, being a local surface term,
namely the local divergence of the Chern-Simons 3-form,
\begin{equation}
\epsilon^{\mu\nu\rho\sigma}F^a_{\mu\nu}F^a_{\rho\sigma}=
\partial_\mu\left\{\epsilon^{\mu\nu\rho\sigma}
\left[A^a_{\nu}F^a_{\rho\sigma}+ 
\frac{1}{3}f^{abc}A^a_\nu A^b_\rho A^c_\sigma\right]\right\}\ .
\end{equation}
Nevertheless, in the presence of a nontrivial spacetime topology,
and especially at the quantum level, such a topological contribution
to the action has important physical consequences for
gauge field configurations of nontrivial topology. For instance,
it suffices to mention in this context the role of instanton
configurations in an euclidean spacetime formulation of the theory,
directly connected to the possibility of this topological term
which, as a matter of fact, also measures the instanton winding number, 
indeed a topological invariant.

In other spacetime dimensions, besides the $F^a_{\mu\nu}F^{a\mu\nu}$ 
contribution to the above Lagrangian density common to all cases, 
analogous $\theta$-topological terms are also possible. In 1+1 dimensions, 
one may consider
\begin{equation}
\frac{1}{2}\theta\,{\rm Tr}\,\epsilon^{\mu\nu}F^a_{\mu\nu}T^a\ ,
\end{equation}
leading to a nonvanishing contribution for gauge groups $G$ including
U(1) factors. In 2+1 dimensions, one may add the Chern-Simons density itself,
\begin{equation}
\frac{1}{2}\theta\epsilon^{\mu\nu\rho}\left[A^a_\mu F^a_{\nu\rho}+
\frac{1}{3}f^{abc}A^a_\mu A^b_\nu A^c_\rho\right]\ .
\end{equation}
And similarly in still higher dimensions of spacetime.

Note that in the above definition of an action principle, and
in comparison with the usual expression for the field strength and
its squared-contribution $F^a_{\mu\nu}F^{a\mu\nu}$ in the action, 
one has performed a rescaling of the gauge field $A^a_\mu$ by the gauge 
coupling constant $e$, namely $A^a_\mu=e\tilde{A}^a_\mu$ with 
$\tilde{A}^a_\mu$ denoting the usual field parametrisation used in 
ordinary perturbation theory. Such a rescaling is quite natural in 
the context of a pure YM theory in the absence of any coupling to 
some matter fields. In terms of the specific limits in the gauge 
coupling constant to be discussed presently, such a rescaling is 
a nontrivial feature, which enables one to explore nonperturbative 
aspects of this dynamics from a potentially novel point of view.

For instance, considering the 4-dimensional action (\ref{eq:YM4}),
it appears obvious that the limit $e\rightarrow 0$ requires one
to restrict to the sector of field configurations of vanishing
field strength,
\begin{equation}
e\rightarrow 0\ :\ \ \ F^a_{\mu\nu}=0\ .
\label{eq:F=0}
\end{equation}
The classification of solutions to these equations is topological
in character, with a set of configurations directly dependent on the
choice of space(time) topology. In the case of an abelian gauge group
for instance, such configurations may be thought of as corresponding 
to Aharonov-Bohm magnetic flux lines threading the different 1-cycles 
associated to the $d$-dimensional spatial torus. In other words, 
the limit of a vanishing gauge coupling constant should lead to a 
topological field theory, as will indeed be made more explicit hereafter, 
showing that the ensuing theory is a so-called TQFT of BF type.\cite{TQFT}
However, note that in terms of the perturbative field $\tilde{A}^A_\mu$, 
given configurations of finite $A^a_\mu$ values the limit $e\rightarrow 0$ 
corresponds to a limit of arbitrarily large values for $\tilde{A}^a_\mu$, 
namely a nonperturbative regime in terms of the usual parametrisation of field
configurations at the basis of ordinary perturbation theory.

Likewise, the other extreme limit of an infinite gauge coupling
constant, $e\rightarrow\infty$, leads to a purely topological field
theory of which the action principle is metric independent and
purely of topological character, reducing to the $\theta$-topological
contribution. In the case of the 4-dimensional
theory (\ref{eq:YM4}), one is simply left with the
instanton density, precisely the system used by Witten as the
TQFT related to Donaldon's invariants for 4-dimensional differentiable
manifolds, leading to (anti)self-dual instanton configurations of
the associated YM theory.\cite{Witten1.1,Baulieu} Note again that the limit 
$e\rightarrow\infty$ requires vanishing $\tilde{A}^a_\mu$ configurations
in order to maintain finite the $A^a_\mu$ ones.
For spacetimes of other dimensions, in a similar manner, the limit
$e\rightarrow\infty$ leads to the associated $\theta$-topological density,
multiplied by a free parameter $\theta$. However, in the topological
limit, often this free parameter requires to take on a quantised
value in order to maintain a nonempty set of gauge invariant physical
states when proper account is given of the large components of the group
of local gauge transformations, the so-called modular group.

Hence, the two extreme limits in the gauge coupling constant lead
indeed to TQFT's of interest and relevance. In the physical sectors
of gauge invariant quantum states of definite mass and energy, one should
also expect that with either limit, some of these states become infinitely
massive and thus decouple, whereas the remainder become massless.
Indeed, in either limit leading to a TQFT, the associated Hamiltonian must 
vanish identically for gauge invariant states, since TQFT's are also 
reparametrisation invariant theories in the spacetime coordinates --- being 
purely topological in the gauge invariant sector --- so that states should also
be independent of the time coordinate along which dynamical evolution is
generated by the Hamiltonian. As is well known, typically, monopole masses
vary according to the inverse of the gauge coupling constant squared, so 
that in either limit indeed one should expect that some states decouple 
by acquiring an infinite mass, while others survive by becoming perfectly 
massless, both types of behaviours being dual to one another through each 
of the two possible limits. Such a dual situation is encountered with
YM theories possessing further supersymmetries, in which the roles of
monopoles and instantons in the confinement and chiral symmetry breaking
phenomena have been unravelled in an analytic way.\cite{Seiberg1,Seiberg2}

This expectation thus also suggests an alternative to ordinary perturbation 
theory based on the trivial field configuration $\tilde{A}^a_\mu=0$. Indeed, 
starting from the sector of gauge invariant states in either of the 
two topological limits, which are thus nonperturbative in character, 
one may turn on either the coupling constant $e$ or its inverse $1/e$ 
and start constructing a perturbation theory based on nonperturbative 
background configurations which reduce to those of a TQFT in the associated 
limit. Hopefully, the two limits could then possibly be merged at 
intermediate coupling values, shedding new light, in a nonperturbative 
manner, on the dynamical issues raised by the nonperturbative phenomena 
at the origin of confinement and chiral symmetry breaking in strongly 
interaction theories such as quantum chromodynamics (QCD). 
Note well than such a procedure relies
from the outset on a sector of quantum states which is gauge invariant
and physical, in contradistinction to ordinary perturbation theory
based on the non gauge invariant states of the gluon and quark field
degrees of freedom.

\vspace{-10pt}

\subsection{The Hamiltonian First-Order Formulation}
\label{Sec2.2}

As made explicit above, the limit $e\rightarrow\infty$ of pure YM theory
leads to a TQFT of a type which is dependent on the spacetime dimension.
In 1+1 dimensions, the Lagrangian density reduces purely to the
abelian field strength $\theta\epsilon^{\mu\nu}F_{\mu\nu}$. In 2+1 dimensions,
it reduces to a pure Chern-Simons dynamics of which the physics resides
purely in the gauge field zero-modes.\cite{Witten2,GovCS} 
In 3+1 dimensions, it should reduce to fields related to some 
(anti)self-dual instanton configurations.\cite{Witten1.1}
Hence, the limit $e\rightarrow\infty$ must be explored for each of these
dimensions separately.

On the other hand, in order to confirm that the limit $e\rightarrow 0$
leads to a TQFT of BF type, it is preferable to turn to the
Hamiltonian formulation, ignoring for the time being the
possible $\theta$-topological contribution proportional to the 
parameter $\theta$. This is a standard exercise in constrained 
dynamics,\cite{GovBook,Gov1}
the outcome of which is as follows, once the physical time coordinate 
$x^0=ct$ is taken to be the evolution variable for the dynamics.

The time component of the gauge field, $A^a_0(t,\vec{x}\,)$, is the
Lagrange multiplier for the generators of small gauge transformations,
themselves represented through first-class constraints expressing
Gauss' law,
\begin{equation}
\phi^a=D_i\pi^{ai}=\partial_i\pi^{ai}-f^{abc}A^b_i\pi^{ci}=0\ ,
\end{equation}
where $\pi^{ai}(ct,\vec{x}\,)$ are the momenta conjugate to the space 
components $A^a_i(ct,\vec{x}\,)$ of the gauge field, obeying the
canonical Poisson brackets,
\begin{equation}
\left\{A^a_i(ct,\vec{x}\,)\,,\,\pi^{bj}(ct,\vec{y}\,)\right\}=
\delta^{ab}\delta^j_i\,\delta^{(d)}(\vec{x}-\vec{y}\,)\ ,
\end{equation}
and $D_i$ denoting the usual gauge covariant derivative, defined here
to act in the adjoint representation of the gauge group $G$.
Up to some factor, $\pi^{ai}$ defines the chromoelectric field.
The first-order (gauge invariant) Hamiltonian density then writes as
\begin{equation}
\mathcal{H}=\frac{1}{2}e^2\pi^{ai}\,\pi^{ai}\ +\
\frac{1}{4e^2}F^a_{ij}F^{aij}\ ,
\end{equation}
so that finally the first-order Lagrangian density of the Hamiltonian
formulation of the system is given as
\begin{equation}
\mathcal{L}_1=\partial_0A^a_i\,\pi^{ai}\ -\
\frac{1}{2}e^2\pi^{ai}\pi^{ai}\,-\,
\frac{1}{4e^2}F^a_{ij}F^{aij}\,+\,A^a_0\left(D_i\pi^{ai}\right)\ .
\label{eq:L1}
\end{equation}

{}From this point of view, the conjugate momenta fields $\pi^{ai}$ appear
simply as auxiliary gaussian fields for which the equations of motion read
\begin{equation}
\pi^{ai}=\frac{1}{e^2}F^a_{0i}\ .
\end{equation}
By direct substitution back in (\ref{eq:L1}), one recovers the
original Lagrangian action of the system. However, this remark may
also suggest to introduce further auxiliary gaussian fields this time
associated to the chromomagnetic sector 
$F^a_{ij}=\partial_iA^a_j-\partial_jA^a_i-f^{abc}A^b_iA^c_j$, in order
to linearise the chromomagnetic energy contribution to the Hamiltonian
density, namely
\begin{equation}
\frac{1}{4}e^2\phi^a_{ij}\phi^{aij}+\frac{1}{2}\phi^a_{ij}F^a_{ij}=
\frac{1}{4}e^2\left(\phi^a_{ij}+\frac{1}{e^2}F^a_{ij}\right)^2-
\frac{1}{4e^2}F^a_{ij}F^{aij}\ ,
\end{equation}
thus with the equation of motion
\begin{equation}
\phi^a_{ij}=-\frac{1}{e^2}F^a_{ij}\ .
\end{equation}
Consequently, combining the fields $\pi^{ai}$ and $\phi^a_{ij}$ into
a covariant tensor $\phi^{a\mu\nu}$ transforming in the adjoint representation
of the gauge group, with the mixed time-space components
$\phi^{a0i}=\pi^{ai}$, the complete Hamiltonian first-order action acquires 
the manifestly covariant form
\begin{equation}
\mathcal{L}_2=\frac{1}{2}\phi^{a\mu\nu}F^a_{\mu\nu}+
\frac{1}{4}e^2\phi^a_{\mu\nu}\phi^{a\mu\nu}\ ,
\end{equation}
where in the second contribution on the r.h.s. the spacetime metric is
to be used to lower the spacetime indices of the antisymmetric field 
$\phi^{a\mu\nu}$.

Finally, performing a spacetime duality transformation exchanging the
chromo-electric and -magnetic sectors of the dynamics,
\begin{equation}
\phi^{a\mu\nu}=\frac{1}{(d-1)!}\epsilon^{\mu\nu\mu_1\mu_2\cdots\mu_{d-1}}\,
\phi^a_{\mu_1\mu_2\cdots\mu_{d-1}}\ ,
\end{equation}
in terms of a totally antisymmetric tensor field 
$\phi^a_{\mu_1\mu_2\cdots\mu_{d-1}}$ transforming in the adjoint representation
of the gauge group, the original YM action is equivalent to the following
first-order action density in covariant Hamiltonian form,
\begin{equation}
\begin{array}{r c l}
\mathcal{L}_3&=&\frac{1}{2!(d-1)!}
\epsilon^{\mu_1\mu_2\cdots\mu_{d-1}\mu\nu}
\phi^a_{\mu_1\mu_2\cdots\mu_{d-1}}\,F^a_{\mu\nu}\ +\\
 & & \\
&&+\frac{(-1)^d}{2!(d-1)!}e^2\phi^a_{\mu_1\mu_2\cdots\mu_{d-1}}
\phi^{a\mu_1\mu_2\cdots\mu_{d-1}}\ .
\end{array}
\label{eq:L3}
\end{equation}
It should be emphasized that in this expression, the fields $A^a_\mu$,
in terms of which $F^a_{\mu\nu}$ is defined, and 
$\phi^a_{\mu_1\cdots\mu_{d-1}}$ are independent. However, it is clear
that the latter field is an auxiliary gaussian one. The variational principle
sets this field proportional to the spacetime dual of the field strength
$F^a_{\mu\nu}$, whereupon substitution into the Lagrangian density
$\mathcal{L}_3$ one immediately recovers the original YM Lagrangian density.

However, the form (\ref{eq:L3}) of the dynamics makes a few interesting
facts explicit. Note that the first contribution in the r.h.s. of
(\ref{eq:L3}) does not involve the spacetime metric structure. This term
is purely topological in character, and indeed defines the TQFT of so-called
BF type (the $B$ field being in this instance the 
$\phi^a_{\mu_1\cdots\mu_{d-1}}$ one).\cite{BFQCD} 
The second contribution on the r.h.s. of (\ref{eq:L3}) combines
both the gauge coupling constant $e$ with the spacetime metric required
to raise the spacetime indices of the antisymmetric field
$\phi^a_{\mu_1\cdots\mu_{d-1}}$.\footnote{Had one from the outset coupled the
YM action to a curved spacetime metric, the same remark would have been
of application.} Again, this very fact hints at the
possibility that geometry --- namely, the essence of the gravitational
interaction --- acquires its full physical meaning only in the presence
of the other (gauge) interactions, whereas a theory of pure quantum gravity
need not necessarily be a quantum dynamics of geometry but possibly 
only a theory of quantum topology, 
namely a TQFT.\cite{Witten1.1,Witten1.2,Witten3}
Finally, note that by having introduced the field 
$\phi^a_{\mu_1\cdots\mu_{d-1}}$ dual to 
the field strength $F^a_{\mu\nu}$, the limit $e\rightarrow 0$ is now
well defined, and indeed reduces solely to the BF TQFT term,\cite{BFQCD} 
confirming the previous expectation based on a formal argument. In particular,
in that limit, the equations of motion of the BF theory are such that
\begin{equation}
F^a_{\mu\nu}=0\ \ \ ,\ \ \ 
\epsilon^{\mu\nu\mu_1\mu_2\cdots\mu_{d-1}}\,
D_\nu\phi^a_{\mu_1\mu_2\cdots\mu_{d-1}}=0\ ,
\end{equation}
thus including the condition of vanishing field strengths in the gauge sector.

\vspace{-5pt}

\subsection{The Topological Quantum Field Theory Limits}
\label{Sec2.3}

Given the above considerations, the two topological limits
of YM dynamics identifying TQFT sectors are best characterised
as follows. For what concerns the $e\rightarrow\infty$ limit,
the original Lagrangian density formulation is best suited,
showing that only the $\theta$-topological contribution
related to Pontryagin/Chern-Simons densities of given order, depending
on the spacetime dimension, survives in this limit, generally also
with a further quantisation restriction on the normalisation
factor of such topological actions, dependent on the spacetime topology
itself.

For what concerns the $e\rightarrow 0$ limit, the first-order
Hamiltonian formulation is better suited, in terms of the original
gauge field $A^a_\mu$ and its field strength $F^a_{\mu\nu}$ together
with the $(D-2)$-form $\phi^a_{\mu_1\cdots\mu_{D-2}}$ transforming
linearly in the adjoint representation of the gauge group. The
topological limit then leads to a TQFT of BF type.

Each of these two possible limits thus identifies a topological
sector of gauge invariant physical states of YM dynamics. Possibly,
these states could serve as a starting point for a manifestly gauge
invariant perturbation expansion of the full YM Hamiltonian above
a nonperturbative set of physical states. At this stage, we leave this
question open, as a possible avenue for further investigation of such
an approach to the nonperturbative dynamics even of pure YM theories.

\vspace{-5pt}

\section{Generalised Pure Yang--Mills Dynamics}
\label{Sec3}

Besides the original motivation for the present programme, the
above rewriting of pure YM dynamics in first-order form suggests
possible generalisations of the usual YM theories, which may well
deserve further study, at least in the cases of 2+1 and 3+1 dimensions.

Let us first consider the case of 1+1 dimensions. The antisymmetric
field then reduces to a single scalar field in the adjoint representation,
so that the first-order Lagrangian density reads,
\begin{equation}
\mathcal{L}^{1+1}_{\rm total}=\frac{1}{2}\epsilon^{\mu\nu}\phi^a F^a_{\mu\nu}
-\frac{1}{2}e^2\phi^a\phi^a+
\frac{1}{2}N\epsilon^{\mu\nu}{\rm Tr}\,F^a_{\mu\nu}T^a\ .
\end{equation}
Clearly, the quadratic invariant term $\phi^a\phi^a$ may be extended to
an arbitrary $G$-invariant potential for the field $\phi^a$, hence
leading to generalised pure YM dynamics in 1+1 dimensions,\cite{GYM2}
\begin{equation}
\mathcal{L}^{1+1}_{\rm total}=\frac{1}{2}\epsilon^{\mu\nu}\phi^a F^a_{\mu\nu}
-V(\phi^a)+\frac{1}{2}N\epsilon^{\mu\nu}{\rm Tr}\,F^a_{\mu\nu}T^a\ .
\end{equation}
It is only in the case of a quadratic potential, associated to the
quadratic Casimir invariant of the gauge group $G$, that the field $\phi^a$
is gaussian and may easily be integrated out. Clearly, for other choices
of the potential contribution, the physics of the system is different,
as may be seen already from a different energy eigenspectrum. Note also
that by extending these considerations to noncompact gauge groups,\cite{Cham}
one could possibly design theories for pure quantum
gravity in 1+1 dimensions possessing a topological sector. Let us
also recall that pure YM dynamics in 1+1 dimensions is an 
example of a totally integrable dynamics,\cite{Witten4} which has not 
yet yielded all its riches though.\cite{deHaro}

With the situation in 2+1 dimensions, there appear already further
possibilities, of which the consequences may well be worth exploring.
The antisymmetric field is then a vector field $\phi^a_\mu$, hence
with the YM dynamics,
\begin{equation}
\mathcal{L}^{2+1}_{\rm total}=
\frac{1}{2}\epsilon^{\mu\nu\rho}\phi^a_\mu F^a_{\nu\rho}+
\frac{1}{2}e^2\phi^a_\mu\phi^{a\mu}+
\frac{1}{2}N\epsilon^{\mu\nu\rho}A^a_\mu\left[F^a_{\nu\rho}+
\frac{1}{3}f^{abc}A^b_\nu A^c_\rho\right]\ .
\end{equation}
Further topological terms may then be added to this action, such as
\begin{equation}
\epsilon^{\mu\nu\rho}{\rm Tr}\,\phi_\mu\phi_\nu\phi_\rho\ \ ,\ \ 
\epsilon^{\mu\nu\rho}{\rm Tr}\,\phi_\mu D_\nu\phi_\rho\ ,
\end{equation}
while the metric-dependent quadratic contribution in the vector field
$\phi^a_\mu$ may be generalised to an arbitrary $G$-invariant potential
$V(\phi^a_\mu\phi^{a\mu})$.

As a matter of fact, the case of pure gravity in 2+1 dimensions is
given by the sole $\theta$-topological action, namely the Chern-Simons
density, for a specific noncompact gauge group depending on the sign of
the cosmological constant.\cite{Witten3} The above generalised action 
could thus provide extensions to the quantum dynamics of pure 
quantum gravity in 2+1 dimensions, a topic in itself worth exploring 
including its eventual physical interpretation.

Finally, one has the situation in 3+1 dimensions, with the antisymmetric
field $\phi^a_{\mu\nu}$ in the adjoint $G$-representation. The original
action reads, in first-order form,
\begin{equation}
\mathcal{L}^{3+1}_{\rm total}=\frac{1}{4}\epsilon^{\mu\nu\rho\sigma}
\phi^a_{\mu\nu} F^a_{\rho\sigma}-\frac{1}{4}e^2\phi^a_{\mu\nu}\phi^{a\mu\nu}+
\frac{1}{4}N\epsilon^{\mu\nu\rho\sigma}F^a_{\mu\nu}F^a_{\rho\sigma}\ .
\end{equation}
Generalisations of this dynamics include the possibility to add further
topological terms, such as
\begin{equation}
\epsilon^{\mu\nu\rho\sigma}\phi^a_{\mu\nu}\phi^a_{\rho\sigma}\ ,
\end{equation}
as well as metric-dependent terms extending the quadratic contribution
of the antisymmetric field, such as
\begin{equation}
\phi^a_{\mu\nu}F^{a\mu\nu}\ \ ,\ \
F^a_{\mu\nu}F^{a\mu\nu}\ \ ,\ \ 
V(\phi^a_{\mu\nu}\phi^{a\mu\nu})\ .
\end{equation}

Such possibilities may well be of interest from the point of view
of particle phenomenology. Indeed, such generalisations of pure YM dynamics
extend the dynamics of the gluons of QCD with further interactions and
contributions accompanied by their own coupling parameters. Hence,
such generalised YM theories provide generalised QCD theories amenable
to experimental tests. In particular, the whole programme of renormalisation
theory, including a calculation of the $\beta$-functions for the relevant
coupling constants and operators including the gauge coupling constant, 
and how they could possibly affect quark and gluon distribution functions 
in hadrons and jet production distributions, could then be subjected 
to experimental tests and constraints, helping in narrowing down 
the actual relevance of ordinary QCD dynamics to the world of particle 
physics. One of the uneasy aspects to QCD is that there is no serious
contender in competition to it, against which to assess the merits
and discriminate the differences and difficulties of each theory,
in confrontation with the experimental facts. With the above generalised
YM dynamics in 3+1 dimensions may well reside such a possibility, thus worth
exploring further.

As a last remark, one may wonder how to define similar considerations
for quantum mechanical systems, namely theories defined in 0+1 dimensions.
The most efficient procedure is to consider the dimensional reduction from
1+1 dimensions to 0+1 dimensions, namely consider the 1+1 dimensional case
with field degrees of freedom of which the space dependence is turned off.
Consequently, one has the degrees of freedom $\phi^a(t)$ together with
coordinates $q^a(t)$ corresponding to the space component of the original
gauge field $A^a_\mu(t,x)$ taken to be $x$-independent, for which 
the dynamics is governed by the Lagrange function,
\begin{equation}
L^{0+1}_{\rm total}=\phi^a D_tq^a\,-\,\frac{1}{2}e^2\phi^a\phi^a+
N{\rm Tr}\,D_tq^aT^a\ ,
\label{eq:0+1}
\end{equation}
where $D_t$ denotes the usual gauge covariant derivative in the time
direction, $D_t\cdot=\partial_t+i[A^a_0(t)T^a,\cdot]$, in the adjoint
$G$-representation. Note that here as well the term quadratic in $\phi^a$ 
may be generalised to an arbitrary $G$-invariant potential $V(\phi^a)$.
Furthermore, for reasons detailed hereafter in the abelian case,
the coordinates $q^a(t)$ are restricted to a specific torus topology
depending on the considered gauge group, as a consequence of large
gauge transformations in the original 1+1 dimensional theory. Finally,
in 0+1 dimensions, an arbitrary $G$-invariant potential in the
coordinates $q^a(t)$ may also be added to the above 
action.\cite{Klauder,Shabanov}

For the sake of illustration, as simple examples of topological
theories we shall consider hereafter systems of the above type, in limits
equivalent to the limits discussed above in terms of the gauge coupling
constant~$e$. Furthermore, we shall restrict to the abelian U(1) gauge
group only with a compact topology.

\vspace{-5pt}

\section{The Abelian Case in 0+1 Dimensions}
\label{Sec4}

As the simplest illustration of the general discussion,
let us turn to the case of a free particle of mass $m$ moving
on a circle of radius $R$, described by the following Lagrange function,
\begin{equation}
L=\frac{1}{2}m\dot{q}^2+N\dot{q}\ ,
\label{eq:particle}
\end{equation}
where $q(t)$ is the particle coordinate, with values identified modulo
integer multiples of $L=2\pi R$, and $N$ an arbitrary constant. 
The second term on the
r.h.s. of this expression is a total time derivative, {\it i.e.\/},
a $\theta$-topological contribution, and thus does not
affect the classical equations of motion. However at the quantum level,
because of the nontrivial circle topology of the configuration space,
the presence of such a term leads to specific physical consequences.
Furthermore, in the limit that $m=0$, even though the Lagrange reduces
to a simple total time derivative, hence leading to a trivial
equation of motion, $0=0$, still at the quantum level one has a nontrivial
quantum situation, albeit then a purely topological quantum system.
Finally, note that the arbitrary normalisation constant $N$ of the
topological term $N\dot{q}$ is physically tantamount to having a constant
U(1) vector potential $A(q)$ present on the circle and to which the
particle would be magnetically coupled.

In terms of the systems discussed previously, it may readily be seen
that (\ref{eq:particle}) is equivalent to the dynamics described by
(\ref{eq:0+1}) in the case of the abelian U(1) gauge group, and once the
auxiliary degree of freedom $\phi(t)$ is integrated out, with the mass
parameter $m$ corresponding to the inverse gauge coupling constant
squared, $m=1/e^2$. The restriction to the circle topology for 
the coordinate $q(t)$ stems from the dimensional
reduction of the compact U(1) gauge field $A_\mu$, as will become clear
in the next section. Indeed, the modular group or group of large U(1)
gauge transformations effectively restricts the configuration space of
the constant mode of the space component of the gauge field --- indeed the
coordinate $q(t)$ --- to the circle topology.

The first-order Hamiltonian formulation of the system is straightforward.
The associated action principle writes as
\begin{equation}
S[q,p]=\int\,dt\,\left[\dot{q}p-\frac{1}{2m}\left(p-N\right)^2\right]\ ,
\end{equation}
which is indeed equivalent in form to (\ref{eq:0+1}). 
Here, $p(t)$ denotes the momentum conjugate to $q(t)$, with the 
canonical Poisson bracket $\{q(t),p(t)\}=1$.

The canonical quantisation of this system is immediate, provided, however,
one takes due account of the circle topology of the configuration space
$q(t)$. As discussed in Refs.~\refcite{Victor1} and \refcite{Victor2}, 
unitarily inequivalent
representations of the Heisenberg algebra $[\hat{q},\hat{p}]=i\hbar$
on the circle are distinguished by a U(1) holonomy parameter $\lambda$ 
defined modulo the integers, $0\le\lambda< 1$, contributing to 
the $\hat{p}$-eigenspectrum which itself is discrete given the compact 
configuration space in $q(t)$. Hence, a basis of Hilbert space is spanned 
by all orthonormalised $\hat{p}$-eigenstates $|n\rangle$ ($n\in\mathbb Z$),
\begin{equation}
\hat{p}|n\rangle = \frac{\hbar}{R}\left(n+\lambda\right)\ \ ,\ \ 
\langle n|m\rangle = \delta_{nm}\ .
\end{equation}
In the configuration space representation with $0\le q<2\pi R$, one has,
\begin{equation}
\begin{array}{r c l}
\hat{q}|q\rangle &=& q|q\rangle\ \ ,\ \ 
\langle q|q'\rangle = \delta(q-q')\ \ ,\ \
\langle q|n\rangle=\frac{1}{\sqrt{2\pi R}}e^{inq/R}\ ,\\
 & & \\
\langle q|\hat{q}|\psi\rangle &=& q\langle q|\psi\rangle\ \ ,\ \ 
\langle q|\hat{p}|\psi\rangle=
\left[-i\hbar\frac{d}{dq}+\frac{\hbar}{R}\lambda\right]\langle q|\psi\rangle\ .
\end{array}
\end{equation}
Consequently, the energy spectrum is given by
\begin{equation}
\hat{H}|n\rangle =\frac{1}{2m}\left[\hat{p}-N\right]^2|n\rangle = E_n|n\rangle
\ \ \ ,\ \ \ 
E_n=\frac{\hbar^2}{2mR^2}\left[\left(n+\lambda\right)-
\frac{R}{\hbar}N\right]^2\ .
\label{eq:En}
\end{equation}
Note the spectral-flow properties of this parabolic spectrum as a function
of the holonomy parameter $\lambda$ defined modulo integer values. Furthermore,
this parameter combines with the $\theta$-topological term normalisation 
constant $N$ into a unique contribution. Both quantities, even though 
combined in this manner, thus indeed have physical consequences at 
the quantum level,
as they, for instance, affect the physical spectrum of the system. This feature
is of purely topological origin, and is made possible by the circle topology
of the $q$-configuration space. All the nontrivial effects of the circle
topology of configuration space contribute through both the holonomy
parameter $\lambda$ and the $\theta$-parameter $N$ combined in the
above fashion. This is a general feature which survives for systems in
higher dimensions. In contradistinction in the decompactification
limit $R\rightarrow\infty$, the spectrum becomes independent of both
$\lambda$ and $N$, as indeed it should since the Heisenberg algebra on
the real line possesses only a single representation, leading back to the
usual system of a free particle on the real line.

The above results may also be used to consider the limits $e\rightarrow 0$
and $e\rightarrow\infty$ discussed previously, this time in terms of the
mass parameter $m$ which determines the parabolic curvature of the parabolic
energy spectrum (\ref{eq:En}). In the limit that $m\rightarrow\infty$, or
$e\rightarrow 0$, the parabola degenerates into a flat spectrum, leaving over
all the states in Hilbert space, but with a vanishing energy. This is the
static limit of the particle model.

The limit $m\rightarrow 0$ or $e\rightarrow\infty$ is that of a purely
topological particle on the circle. All states acquire then an infinite
energy and decouple, except at most for one state $|n_0\rangle$ provide
the combination of constants $N$ and $\lambda$ (which could both depend on
$m$) is chosen such that
\begin{equation}
\lim_{m\rightarrow 0}N(m)=N_0=\frac{\hbar}{R}(n_0+\lambda_0)\ .
\label{eq:N0}
\end{equation}
In other words, the normalisation of the action for the topological
particle needs to be quantised according to this rule, in order to
obtain a nonempty spectrum of states. In the present case, this
spectrum of states is one dimensional, which is a characteristic
feature of TQFT's. Indeed, generally even in quantum mechanics
(think of the harmonic oscillator for instance), the space of quantum
states is infinite dimensional, albeit sometimes discrete, while 
for TQFT's it is often finite dimensional, meaning that the physical classical
phase space itself is then also compact (which is indeed the case 
for our example in the limit $m=0$).

As a matter of fact, the topological particle is a constrained 
system,\cite{GovBook,Gov1} with the constraint
\begin{equation}
p(t)-N=0\ .
\end{equation}
The gauge invariance associated to this constraint is that of
arbitrary ``field'' redefinitions of the coordinate $q(t)$. Quantum
physical states should thus be annihilated by this constraint, hence
for the present case indeed leading to a single state in Hilbert
space, provided the normalisation constant $N$ obeys the quantisation
condition (\ref{eq:N0}). Note however that these features directly
related to the nontrivial topology of configuration space, and implying
a nontrivial physical content associated to an action which is a pure total
derivative, are totally ``dissolved'' away in the decompactification limit.
Indeed, the normalisation factor $N$ then needs to vanish altogether,
leaving over a trivially vanishing action for the system. The topology
of the real line being trivial, the TQFT of the topological particle
does not provide any information. However, for a circle topology,
the topological particle leads to a nontrivial quantum system.
As a matter of fact, which shall not be detailed further here,
given the circle topology, namely precisely the U(1) group manifold,
all the possible quantum realisations of the topological particle
are in one-to-one correspondence with the irreducible representations
of the U(1) group, indeed labelled by the integers $n_0$ specifying
the single physical state surviving in the TQFT limit $m=0$. The
same conclusion applies to the topological particle defined over general
compact Lie group manifolds and their compact cosets.\cite{Jarvis}

\vspace{-5pt}

\section{The Abelian Case in 1+1 Dimensions}
\label{Sec5}

As another simple illustration of our general programme,
let us now consider the pure U(1) gauge theory in 1+1 dimensions,
defined by the Lagrangian density
\begin{equation}
\mathcal{L}=-\frac{1}{4e^2}F_{\mu\nu}F^{\mu\nu}+
\frac{1}{2}N\epsilon^{\mu\nu}F_{\mu\nu}\ ,
\end{equation}
or equivalently
\begin{equation}
\mathcal{L}=\frac{1}{2}\epsilon^{\mu\nu}\phi F_{\mu\nu}
-\frac{1}{2}e^2\phi^2+\frac{1}{2}N\epsilon^{\mu\nu}F_{\mu\nu}=
\phi F_{01}-\frac{1}{2}e^2\phi^2+N F_{01}\ ,
\end{equation}
where
\begin{equation}
F_{\mu\nu}=\partial_\mu A_\nu - \partial_\nu A_\mu\ .
\end{equation}
The space coordinate $x$ is of course assumed to be compactified
into a circle $S$ of radius $R$ and circumference $L=2\pi R$.

By construction, this system is invariant under local compact U(1)
gauge transformations, of the form,
\begin{equation}
A'_\mu(t,x)=A_\mu(t,x)+\partial_\mu\theta(t,x)=
A_\mu(t,x)+\partial_\mu\theta_0(t,x)+\sum_{k\in\mathbb Z}
\frac{k}{R}\delta_{\mu,1}\ ,
\end{equation}
where the U(1) phase parameter $\theta(t,x)$ of the transformation includes
both a contribution $\theta_0(t,x)$ periodic on the circle in space, 
associated to small gauge transformations, as well as the large gauge 
transformation component corresponding to the $\mathbb Z$ modular group of
topologically nontrivial U(1) transformations on the circle distinguished
by the winding number $k\in\mathbb Z$,
\begin{equation}
\theta(t,x)=\theta_0(t,x)+\sum_{k\in\mathbb Z}\,k\frac{x}{R}\ \ ,\ \ 
\theta_0(t,x+2\pi R)=\theta_0(t,x)\ ,
\end{equation}
such that $e^{i\theta(t,x)}$ be single-valued on the considered
spacetime topology $\mathbb R\times S$.

In explicit form, the first-order action density reads, after integrating
by parts in $x$ and assuming the fields $A_\mu(t,x)$ and $\phi(t,x)$ to
be $2\pi R$-periodic in $x$,
\begin{equation}
\mathcal{L}=\partial_tA_1(\phi+N)-\frac{1}{2}e^2(\phi+N-N)^2+
A_0\partial_x(\phi+N)\ .
\end{equation}
Consequently, it follows that $A_1(t,x)$ and $\phi(t,x)$ are canonically
conjugate degrees of freedom, with the Poisson brackets
$\{A_1(t,x),\phi(t,y)\}=\delta(x-y)$, while one has the first-class
Hamiltonian density $\mathcal{H}=e^2\phi^2/2$ and first-class
constraint $\partial_x\phi=0$, which generates the small U(1)
gauge transformations. The time component $A_0(t,x)$ appears to be nothing 
but the Lagrange multiplier for the first-class constraint.

Rather than addressing the full-fledged quantisation of this system,
let us consider the short-cut approach by first identifying the gauge invariant
degrees of freedom of the physical sector at the classical level.
Given the gauge transformations in Hamiltonian form
\begin{equation}
A'_0=A_0+\partial_t\theta_0\ \ ,\ \ 
A'_1=A_1+\partial_x\theta_0+\frac{2\pi k}{L}\ \ ,\ \ 
\phi'=\phi\ ,
\end{equation}
it appears that whatever the configurations for $A_0(t,x)$ and $A_1(t,x)$,
it is always possible to find a small gauge transformation $\theta_0(t,x)$
such that $A_1(t,x)$ is reduced to its space zero-mode, independent
of $x$, namely $A_1(t,x)=\bar{A}_1(t)$, while the Lagrange multiplier
may even be set to vanish, $A_0(t,x)=0$. This exhausts the entire freedom
in performing small gauge transformations parametrised by the $x$-periodic
functions $\theta_0(t,x)$, However, under the large gauge transformation
of homotopy class $k$, the $A_1$ zero-mode $\bar{A}_1(t)$ is still
affected, being shifted by a $k$-multiple of $2\pi/L=1/R$. In other words,
invariance of the system under the modular group $\mathbb Z$ of large
gauge transformations implies the circle topology for its physical
configuration space $\bar{A}_1(t)$. Finally
for what concerns the gauge invariant field $\phi(t,x)$, the first-class
constraint of Gauss' law $\partial_x\phi=0$ is solved by restricting
this sector of degrees of freedom also to its space zero-mode
$\bar{\phi}(t)$.

Consequently, at the classical level, the gauge invariant physical
content of the system reduces to that of the zero-modes $L\bar{A}_1(t)$
and $\bar{\phi}(t)$, which are canonically conjugate to one another.
Indeed, upon this reduction to the physical
sector, the first-order Hamiltonian action reads,
\begin{equation}
S=\int dt\,L\left[\partial_t\bar{A}_1\bar{\phi}
-\frac{1}{2}e^2\bar{\phi}^2+N\partial_t\bar{A}_1\right]\ ,
\end{equation}
in which it must be understood that the coordinate $L\bar{A}_1$ is
to take its values in the circle of radius unity. In other words,
the physics of pure U(1) abelian gauge theory in 1+1 dimensions,
including a $\theta$-topological contribution, is that of a massive
particle on a circle including its $\theta$-topological term.
As a matter of fact, this conclusion remains valid even for a pure
nonabelian YM theory in 1+1 dimensions.\cite{Shabanov}

When comparing to the discussion of the previous section, we see
that the identification is such that, for instance, $L\bar{A}_1(t)$ 
corresponds to $q(t)$, $(\bar{\phi}(t)+N)$ corresponds to $p(t)$, while
the radius parameter is now given by $R=1$ with $e^2L$
corresponding to $1/m$. Consequently, the spectrum
of physical states reduces to the complete set of all states
$|n\rangle$ such that
$\hat{\phi}|n\rangle=[\hbar(n+\lambda)-N]|n\rangle$, while the physical
energy spectrum of the pure U(1) gauge theory is given by
\begin{equation}
E_n=\frac{1}{2}e^2L\hbar^2\left[(n+\lambda)-\frac{N}{\hbar}\right]^2\ .
\end{equation}

The $e\rightarrow 0$ limit is thus tantamount to the totally
compactified limit $L\rightarrow 0$, leading to states all of vanishing
energy, corresponding to a BF TQFT whose Hilbert space is spanned by
all momentum eigenstates $|n\rangle$. In the other extreme limit 
$e\rightarrow\infty$, tantamount to the decompactification limit 
$L\rightarrow\infty$, the TQFT with a single quantum physical state
$|n_0\rangle$ is recovered, provided however the $\theta$-topological
contribution to the action is normalised in such a manner that
\begin{equation}
\lim_{e\rightarrow\infty}N(e)=\hbar(n_0+\lambda_0)\ ,
\end{equation}
reproducing again the quantisation condition on this
normalisation factor. Note also that the spatial circle in the
coordinate $x$ of circumference $L$ implies the circular topology
of radius $1/L$ for the modular space of the system defined by the
zero-mode $\bar{A}_1(t)$. Thus, these extreme compactification or
decompactification limits in physical space correspond to the
opposite process in the actual physical configuration space of the
system, namely its modular space spanned by the variable $\bar{A}_1(t)$.
This feature is reminiscent of T-duality in string theory.\cite{Gov1} 
Could it be that in fact our physical space is the modular space of some
unknown Yang--Mills theory living in some yet undiscovered and very
much compactified physical universe?

Note that the above explicit and exact results illustrate
a property expected on general grounds, and mentioned previously.
Namely the fact that when the gauge coupling constant $e$ varies
between $0$ and $\infty$, the spectrum of physical states includes
states that either decouple by acquiring an infinite energy, or else
degenerate to a vanishing energy. For the simplest systems discussed
here, only one of these classes of states exists; there are no quantum 
states acquiring vanishing or infinite energy while some other subset behaves
in the opposite fashion as the gauge coupling constant is varied.
However, starting with systems in 2+1 dimensions, the latter behaviour
is also observed.\cite{Seiberg1,Seiberg2,Payen}

\vspace{-12pt}

\section{Conclusions}
\label{Sec6}

The main purpose of the present communication is to outline a
general programme towards a study of the nonperturbative dynamics
of Yang--Mills theories which could possibly be at the origin of
the challenging problems of confinement and chiral symmetry breaking
in theories such as quantum chromodynamics. 
The approach is motivated by the suggestion
that this dynamics could be dominated by sectors of Yang--Mills
dynamics which to first approximation are those of Topological Quantum
Field Theories. Indeed, topological quantum effects in the configuration
space of these systems have for long been suspected to play an
essential role in these nonperturbative issues, through monopole and instanton
solutions to the Yang--Mills equations. The present approach also
presents the advantage of considering from the outset gauge invariant
physical quantum states only, defining such nonperturbative topological
configurations, in terms of which to set up a novel type of
perturbation theory. The TQFT sectors that could be identified from
the YM dynamics are reached in the two extreme limits of either a vanishing
or an infinite gauge coupling constant.

Some of the features expected on general grounds could also be illustrated
in the case of two simple examples of such systems, explicitly showing
the importance of properly accounting for the topology of physical and
configuration spaces in Yang--Mills dynamics.

In the course of the discussion a series of open and fascinating
problems were also suggested, which could well prove to be worth
exploring further. These problems span from the phenomenology of
the strong interactions and their experimental study to even the quest
for a final unification of all particles and interactions including
a theory for quantum gravity. It is hoped that some participants
to this Workshop will be sufficiently inspired by one or another
of these open problems, to launch their own efforts with a
curiosity-driven enthusiasm, and contribute to their resolution
and to progress in our understanding of the fascinating and
beautiful riches of our physical universe.

\vspace{-8pt}

\section*{Acknowledgements}

The author wishes to thank Robert de Mello Koch and Jo\~ao Rodrigues
(both at the University of the Witwatersrand, South Africa), 
Hendrik Geyer and Frederik Scholtz (both at the University of Stellenbosch, 
South Africa), Peter Jarvis (University of Tasmania, Australia) and
Florian Payen (Catholic University of Louvain, Louvain-la-Neuve, Belgium),
for constructive discussions on the topics of this contribution.
The work of the author is partially supported by the Federal Office 
for Scientific, Technical and Cultural Affairs (Belgium) 
through the Interuniversity Attraction Pole P5/27.


\vspace{-8pt}

\begin{thebibliography}{99}

\bibitem{Witten1.1}
E. Witten, {\it Comm. Math. Phys.\/} {\bf 117}, 353 (1988).

\bibitem{Witten1.2}
E. Witten, {\it Comm. Math. Phys.\/} {\bf 118}, 411 (1988); 
{\it Phys. Lett. B\/}{\bf 206}, 601 (1988);\\
J.M.F. Labastida, M. Pernici and E. Witten, 
{\it Nucl. Phys. B\/}{\bf 310}, 611 (1988).

\bibitem{Witten2}
E. Witten, {\sl Comm. Math. Phys.\/} {\bf 121}, 351 (1988).

\bibitem{TQFT}
For a review, see,\\
D. Birmingham, M. Blau, M. Rakowski and G. Thompson, 
{\it Physics Reports\/} {\bf 209}, 129 (1991).

\bibitem{GovCS}
For further references, see,\\
J. Govaerts and B. Deschepper, {\it J. Phys. A\/}{\bf 33}, 1031 (2000).

\bibitem{Witten3}
E. Witten, {\it Nucl. Phys. B\/}{\bf 311}, 46 (1988);
{\it Phys. Rev. Lett.\/} {\bf 62}, 501 (1989);
{\it Nucl. Phys. B\/}{\bf 323}, 113 (1989).

\bibitem{vanBaal}
For a review and references, see,\\
F. Bruckmann, D. Nogradi and P. van Baal, {\sl Instantons and constituent
monopoles\/}, {\it Acta Phys. Polon. B\/}{\bf 34}, 5717 (2003),
e-print {\tt arXiv:hep-th/0309008} (September 2003).

\bibitem{Wilson}
K. Wilson, {\it Phys. Rev. D\/}{\bf 10}, 2445 (1974).

\bibitem{Mandelstam}
S. Mandelstam, {\it Physics Reports\/} {\bf 23}, 245 (1976);
{\it Phys. Rev. D\/}{\bf 19}, 2391 (1979);
{\it Physics Reports\/} {\bf 67}, 109 (1980).

\bibitem{tHooft}
G. 't Hooft, {\it Nucl. Phys. B\/}{\bf 138}, 1 (1979);
{\it ibid. B\/}{\bf 153}, 141 (1979);
{\it Comm. Math. Phys.\/} {\bf 81}, 267 (1981);
{\it Phys. Scripta\/} {\bf 25}, 133 (1982);
{\sl Monopoles, instantons and confinement\/}, 
e-print {\tt arXiv:hep-th/0010225}
(October 2000).

\bibitem{Polyakov}
A.M. Polyakov, {\it Phys. Lett. B\/}{\bf 59}, 82 (1975);
{\it ibid. B\/}{\bf 72}, 477 (1978);
{\it Nucl. Phys. B\/}{\bf 120}, 429 (1977).

\bibitem{GovBook}
J. Govaerts, {\sl Hamiltonian Quantisation and Constrained
Dynamics\/} (Leuven University Press, Leuven, 1991).

\bibitem{Gov1}
For a discussion, see for example,\\
J. Govaerts, {\sl The quantum geometer's universe: particles,
interactions and topology\/}, in the Proceedings of the
Second International Workshop on Contemporary Problems in
Mathematical Physics, J.~Govaerts, M.N.~Hounkonnou and A.Z.~Msezane, eds.
(World Scientific, Singapore, 2002), pp. 79--212.

\bibitem{Baulieu}
L. Baulieu and I.M. Singer, {\sl Topological Yang--Mills symmetry\/},
in the Proceedings of the Annecy Meeting on Conformal Field Theories
and Related Topics, Annecy (France), March 14--16, 1988, 
P. Binetruy, P. Sorba and R.~Stora, eds., 
{\it Nucl. Phys. Proc. Suppl. B\/}{\bf 5}, 12 (1988).

\bibitem{Seiberg1}
N. Seiberg and E. Witten, {\it Nucl. Phys. B\/}{\bf 426}, 19 (1994);
Erratum, {\it ibid. B\/}{\bf 430}, 485 (1994); 
{\it ibid. B\/}{\bf 431}, 484 (1994);
{\it ibid. B\/}{\bf 435}, 129 (1995);\\
F. Cachazo, N. Seiberg and E. Witten,
{\it JHEP\/} {\bf 0302}, 042 (2003), e-print {\tt arXiv:hep-th/0301006} 
(January 2003); {\it JHEP\/} {\bf 0304}, 018 (2003),
e-print {\tt arXiv:hep-th/0303207} (March 2003);\\
and further references therein.

\bibitem{Seiberg2}
For a review, see,\\
L. Alvarez-Gaume and S.F. Hassan,
{\it Fortsch. Phys.\/} {\bf 45}, 159 (1997),
e-print {\tt arXiv:hep-th/9701069} (January 1997).

\bibitem{BFQCD}
A.S. Cattaneo, P. Cotta-Ramusino, F. Fucito, M. Martellini, 
M. Rinaldi, A.~Tanzini and M. Zeni,
{\it Comm. Math. Phys.\/} {\bf 197}, 571 (1998);\\
A. Accardi, A. Belli, M. Martellini and M. Zeni,
{\it Phys. Lett. B\/}{\bf 431}, 127 (1998);\\
and further references therein.

\bibitem{GYM2}
M.R. Douglas, K. Li and M. Staudacher,
{\it Nucl. Phys. B\/}{\bf 420}, 118 (1994);\\
O. Ganor, J. Sonnenschein and S. Yankielowicz,
{\it Nucl. Phys. B\/}{\bf 434}, 139 (1995);\\
M. Billo, A. D'Adda and P. Provero,
{\it Nucl. Phys. B\/}{\bf 576}, 241 (2000).

\bibitem{Cham}
A.H. Chamseddine and D. Wyler, {\it Phys. Lett. B\/}{\bf 228}, 75 (1989);
{\it Nucl. Phys. B\/}{\bf 340}, 595 (1990).

\bibitem{Witten4}
E. Witten, {\it Comm. Math. Phys.\/} {\bf 141}, 153 (1991);
{\it J. Geom. Phys.\/} {\bf 9}, 303 (1992);\\
and further references therein.

\bibitem{deHaro}
S. de Haro, {\sl Chern-Simons theory in lens spaces from 2$d$
Yang--Mills on the cylinder\/}, e-print {\tt arXiv:hep-th/0407139}
(July 2004);\\
and further references therein.

\bibitem{Victor1}
J. Govaerts and V.M. Villanueva, 
{\it Int. J. Mod. Phys. A\/}{\bf 15}, 4903 (2000).

\bibitem{Victor2}
J. Govaerts, {\sl Quantisation and gauge invariance\/},
in the Proceedings of the First International Workshop 
on Contemporary Problems in Mathematical Physics, 
J.~Govaerts, M.N.~Hounkonnou and W.A. Lester, Jr., eds.
(World Scientific, Singapore, 2000), pp. 244--259.

\bibitem{Klauder}
J. Govaerts and J.R. Klauder,
{\it Ann. Phys.\/} {\bf 274}, 251 (1999).

\bibitem{Shabanov}
S.V. Shabanov, {\it Physics Reports\/} {\bf 325}, 1 (2000).

\bibitem{Jarvis}
J. Govaerts and P.D. Jarvis, work in progress.

\bibitem{Payen}
F. Payen, Master's Degree Diploma Thesis (Catholic University of
Louvain, Louvain-la-Neuve, Belgium), unpublished (August 2004).

\end{thebibliography}
\end{document}